\documentstyle[preprint,aps]{revtex}
\begin{document}
\begin{titlepage}
%\narrowtext
\title{Pion-Nucleon Scattering and the $\pi$NN  Coupling Constant
in the Chiral Color Dielectric Model}

\author{Dinghui Lu,  Shashikant C. Phatak* and Rubin H. Landau}
\address{Department of Physics, Oregon State University,
Corvallis OR 97331, USA}
\footnotetext {$\mbox{}^*$ Permanent address:
Institute of Physics, Sachivalaya Marg, Bhubaneswar INDIA}
\date{\today}
\maketitle

\begin{abstract}
We apply the chiral color dielectric model to low- and medium-energy
scattering within the coupled $\pi$N and $\pi\Delta$ system.  Dynamic
baryon states in which quarks are confined by the scalar
color-dielectric field are constructed in Fock space.  Spurious motion
of the center of mass is eliminated by constructing momentum
eigenstates via a Peierls-Yoccoz projection. The relativistic
Lippmann-Schwinger equation is solved for the complex energies of the
T matrix poles, and  the pole positions of the nucleon and delta are used
to fix the few parameters of the model.  The $S$- and $P$-wave phase
shifts, the bare N and $\Delta$ masses, the renormalized $\pi$NN and
$\pi$N$\Delta$ coupling constants, and the $\pi$NN and $\pi$N$\Delta$
vertex functions are predicted.
\end{abstract}

%\pacs{PACS numbers: 12.39.Fe, 12.39.Ki, 13.75.Gx, 24.10.Eq}
\end{titlepage}

\section{Introduction}

{}From their first introduction, quark models have been extensively used
to study baryon spectra. Some models have the quarks bound by a
potential\cite{pot}, others have them confined to a bag\cite{bag}, and
others have them bound as solitons\cite{Wilets}. These models are
fairly successful at fitting the masses and static properties of the
baryon octet and decuplet and even some higher excited states.  In
most quark models, the baryon masses are identified with the bound
state energies of quarks within some potential well.  This means that
the pion-baryon interactions do not directly affect the values of the
baryon masses, and that the baryons are stable since decay channels
are not coupled in.  Phenomenologically, the masses of excited baryons
are determined as the resonance energies in the phase-shift analysis
of pion-nucleon scattering.  These experimental masses naturally
include mass shifts due to the pion-baryon interactions as well as
widths due to the finite lifetimes of the states. Accordingly, there
are uncertainties in a direct comparison of the experimental masses to
the bound-state energies of quarks bound in a potential well.

Chiral models, such as the cloudy bag models\cite{cbm}, contain pions
in additions to quarks, and therefore also contain elementary
pion-quark interactions. The calculation of masses with these models
consequently include shifts due to the pion-baryon
interactions. However, these models usually use perturbation theory to
calculate the widths of the excited baryons, and this means that
higher-order terms are ignored and that unitarity in the pion-nucleon
channel is not ensured.  Ideally, one should determine the masses and
widths of excited baryons by calculating the pion-baryon scattering
matrix for a given model, and then finding the complex energies at
which the scattering matrix has poles. This approach was used by
Thomas, Th\'{e}berge and Miller\cite{MTT} for the N-$\Delta$ system,
and later extended to other partial waves\cite{triumf} and to other
mesons\cite{ghe}.

Bag models suffer from several limitations. First, the use of an
artificial bag to confine quarks leads to wave functions which have
sharp discontinuities at the bag boundary. This leads to pion-baryon
form factors (and ultimately potentials) which have unphysical
oscillations and, presumably, limited ability to describe the
scattering data. In addition, bag calculations treat the bare baryon
masses and the bag radius as independent parameters even though we know
this is not correct since for massless quarks the baryon masses are
inversely proportional to the bag radius\cite{bag}.

In the present work the chiral color dielectric model (CDM) is
employed to overcome some of the limitations of bag models.  The CDM
was first introduced phenomenologically to simulate the absolute
confinement of quantum chromodynamics (QCD). A connection with the
more fundamental theory was established by Nielson and Patkos\cite{NP}
who found that an effective color dielectric field does indeed arise
naturally in lattice QCD after defining ``coarse grained'' effective
field variables. The color dielectric field takes into account the
long distance behavior of the QCD vacuum and produces a natural
confinement of the quarks within baryons. More recently, Krein et
al.\cite{krein} have solved the Schwinger-Dyson equation and shown how
confinement and a pseudoscalar meson (Goldstone boson) arise if chiral
symmetry is dynamically broken.

In some recent CDM models, chiral symmetry is restored via a nonlinear
realization which introduces an elementary pion field as a Goldstone
boson.  Even though the mesons in these models are elementary, the
baryon structure and the resulting relations of coupling constants are
determined by the quark model\cite{CDM1,williams}.  In application,
the CDM is similar to the cloudy bag model\cite{cbm,williams}, yet
somewhat more microscopic in its dynamic mechanism for quark
confinement and its treatment of baryon recoil within a
nontopological-soliton solution to the field equations. Accordingly,
when in the present paper we compare models which differ mainly in
their treatment of quark binding, we hope to determine how important
are improved (or maybe just different) quark dynamics when calculating
two-body scattering, and how might the differences in the models
manifest themselves.

Our calculation proceeds as follows. First we solve the quark and
dielectric field equations of CDM in the mean field approximation
(MFA). We then quantize the classical dielectric field using the
coherent state, and improve upon the MFA by including one-gluon
exchange corrections perturbatively.  We include the recoil of the
baryon via the prescription of Peierls-Yoccoz projection\cite{PY}.
This, presumably, is a more
important correction to scattering where large momentum transfers can
occur than for static properties where the momentum transfers are
small. Once we have bare baryons constructed from quarks within  a binding
field, we expand the quark-pion interaction in powers of the pion
field and calculate the transition matrix elements for baryon coupling
with multiple pions.  From these we identify the pion-nucleon
potential, and then dress the vertices by substituting the potentials
into the relativistic Lippmann-Schwinger equation\cite{ghe}:
\begin{eqnarray}
	\lefteqn{T_{\beta \alpha}^{LJI}(k',k)
	= V_{\beta \alpha}^{LJI}(k',k)}
	\nonumber \\
	&& \ + \frac{2}{\pi}\sum_\gamma\int_0^\infty\!dp\,
	\frac{p^2 V_{\beta \gamma}^{LJI}(k',p)T_{\gamma \alpha}^{LJI}(p,k)}
	{E + i\epsilon -E_\gamma(k)}. \label{T}
\end{eqnarray}
Here the superscripts $LJI$ indicate the orbital angular momenta, the
total angular momenta, and the total isospin, and the subscripts
$\beta$ and $\alpha$ indicate the $\pi$N or $\pi \Delta$ channels. We
adjust the parameters of color dielectric model so that the poles of T
matrix elements
occur at the nucleon mass and at the $\Delta$'s complex mass. We then use the
model to predict  low- and medium-energy $\pi$N scattering in $S$
and $P$ waves and the $\pi$NN and the $\pi$N$\Delta$ coupling constants.

\section{Chiral Color Dielectric Model}

The CDM Lagrangian we choose to use is:\cite{CDM1}
\begin{eqnarray}
	\protect{\cal L}
	&=&  \overline q \left[ i\gamma^\mu \partial_\mu -
	{m_q \over \chi} e^{i \gamma_5 \vec{\tau} \cdot \vec{\phi}/f}
	- {g \over 2}  \lambda_a \gamma^\mu A^a_\mu \right ] q \nonumber \\
	&&-{\kappa(\chi) \over 4} (F^a_{\mu\nu})^2
	+ {\sigma_v^2 \over 2} (\partial_\mu \chi)^2 - U(\chi)\nonumber \\
	&&+ {(D_\mu \vec{\phi})^2 \over 2}
	- {m^2_\pi \vec{\phi}^2 \over 2}  , \label{L}\\
        \kappa(\chi) &=& \chi^4, \label{K}\\
	D_\mu\vec{\phi} &=& \hat \phi
	\partial_\mu\phi+f\sin{\phi\over f}
	\partial_\mu \hat \phi, \\
	F^{a}_{\mu\nu} &= &
	\partial_\mu A_\nu^a-\partial_\nu A_\mu^a
		- gf_{abc}A_\mu^b A_\nu^c. \label{F}
\end{eqnarray}
Here $q$, $\vec{\phi}$, $A^a_\mu$ and $\chi$ are the quark, pion,
gluon, and dielectric fields respectively, $\kappa(\chi)$ is the
dielectric coefficient (other CDM models may have different functional
forms for $\kappa(\chi)$), $F^{a}_{\mu\nu}$ is the gluon field tensor,
$\vec{\tau}$ is the isospin operator of the pion, $\lambda_a$ and $f_{abc}$
are Gell-Mann matrices and SU(3) structrue constants, $m_q$ and $m_\pi$
are the quark and pion masses, and $D_\mu\vec{\phi}$ is the covariant
derivative of the pion field. Arrows over
quantities denote their isovector nature. The form (\ref{L}) of the
Lagrangian becomes invariant under chiral transformations when the
pion mass is set equal to zero.

An important, yet phenomenological, part of the CDM Lagrangian
(\ref{L}) is the dielectric self-interaction field $U(\chi)$. We
express it in terms of the bag constant $B$, a shape parameter
$\alpha$, and the dielectric field $\chi$:
\begin{equation}
	U(\chi) = B\alpha\chi^2 \left[1-2\left(1-{2\over \alpha}\right)\chi
	+\left(1-{3 \over
	\alpha}\right)\chi^2\right]. \label{U}
\end{equation}
This  self-interaction field has an absolute minimum when
the dielectric field $\chi=0$ and $\alpha > 6$, and has a local minimum when
$\chi=1$.  Of particular importance is the behavior of the Lagrangian
(\ref{L}) in the limit of vanishing dielectric field, $\chi
\rightarrow 0$.   In this limit, the gluon kinetic energy term
$\chi^4 F^2/4$ vanishes and the quark effective mass $m_q/\chi$
becomes infinite.  Accordingly, we have the distinguishing
characteristic of the CDM: the quark and gluon fields are confined
within the baryon to the region of non-vanishing dielectric field
$\chi$. This contrasts with the bag model in which confinement is
externally imposed by placing the quarks within an infinite square
well.

As we see from the form of the Lagrangian (\ref{L}) with the
dielectric self-interaction field (\ref{U}), the CDM has five
parameters: the quark mass $m_q$, the strong coupling constant $g$ (or
equivalently $\alpha_s = g^2/ 4\pi$), the pion decay constant $f$, the
glueball (or dielectric field) mass $m_{gb} \equiv
{\sqrt{2B\alpha}/\sigma_v}$, and the bag constant $B$.  Although we do
adjust some of these parameters to experimental data, their values are
far from arbitrary.  For example, the glueball mass $m_{gb}$
essentially sets the scale of energy, and the values of the coupling
and decay constants must be close to the accepted values for the model to
be considered realistic.

\subsection{Pseudo-vector Coupling}

In order for the Weinberg-Tomozawa relation to appear explicitly in
S-wave $\pi$N scattering at the tree level (and thus guarantee that the Born
approximation scattering lengths are reasonable), we perform a unitary
transformation\cite{cbm} on the quark fields, $q \rightarrow
\exp[i\gamma_5\vec{\tau}\cdot \vec{\phi}/(2f)] q$.  This transforms
the original CDM Lagrangian (\ref{L}) to
\begin{eqnarray}
	\lefteqn{\protect{\cal L}
	= \overline q \left[ i\gamma^\mu D_\mu
	- {m_q \over {\chi}}
	+ {1 \over 2f} \gamma^\mu \gamma_5 \vec{\tau}\cdot D_\mu
	\vec{\phi} - {g \over 2}  \lambda_a \gamma^\mu A^a_\mu \right
	] q }\nonumber \\
	&& -
	{\kappa(\chi) \over 4} (F^a_{\mu\nu})^2 + {{\sigma_v^2 \over 2}}
	(\partial_\mu \chi)^2 - U(\chi) + {(D_\mu \vec{\phi})^2 \over 2}  -
	{ m^2_\pi \vec{\phi}^2 \over 2} , \label{L2}
\end{eqnarray}
where the covariant
derivative on the quark fields are defined as
\begin{equation}
D_\mu q = \partial _\mu q - {i \over 2} \left(\cos{\phi\over f}-1\right)
\vec{\tau}\cdot \hat \phi \times \partial _\mu \hat \phi q.
\end{equation}
These new quark fields are  dressed by the pions. We next expand
the Lagrangian in powers of the inverse of
$f$, and retain terms up to order $1/f^2$. We separate off  terms
linear and quadratic in $1/f$ and identify them as the interaction
Hamiltonian:
\begin{equation}
	H_{int}  \simeq H_{1\pi} + H_{2\pi},
\end{equation}
where
\begin{eqnarray}
	H_{1\pi} &=& -{1\over 2f} \int d^3 x \, \overline q
	\gamma^\mu\gamma_5 \vec{\tau} q\cdot \partial_\mu \vec{\phi}, \\
	 H_{2\pi} &=& {1\over 4f^2} \int d^3 x \, \overline q
	\gamma^\mu \vec{\tau} q \cdot
	\vec{\phi} \times \partial_\mu \vec{\phi}.  \label{H}
\end{eqnarray}
These elementary vertices are illustrated in Figure~\ref{vertices}.
You will notice that the one-pion vertex $H_{1\pi}$ has a
pseudovector coupling between the pion and the quark.  The only
surviving term for S-wave scattering is the two-pion contact
interaction $H_{2\pi}$ which comes from the covariant derivative on
quarks; this term reproduces the Weinberg-Tomozawa result which was
originally derived from current algebra.

Both the scalar dielectric field and the screened gluon fields exist in
the Lagrangian (\ref{L2}). The simplest treatment of them is in a
mean field approximation in which we ignore the screened gluon fields and the
pion clouds (one then needs one-gluon-exchange ``corrections'' to
split the nucleon and delta masses).  Pions are included
perturbatively in the theory as we derive the bare vertex function and
later dress the bare  vertices by means of the Lippmann-Schwinger
equation (\ref{T}).

\subsection{Vertex Functions with the  Static Baryons}

The equations of motion for the static quark and
dielectric fields in MFA are:
\begin{eqnarray}
 	\left(i\gamma^\mu \partial _\mu - {m_q \over {\chi}} \right) q
	&=& 0, \label{field1}\\
	\sigma_v^2 \partial_\mu \partial^\mu \chi
	+ {dU(\chi)\over d\chi} -{m_q \over
	\chi^2}  \, \overline q q &=& 0.\label{field2}
\end{eqnarray}
These are coupled, nonlinear, partial differential equations.  We
solve them numerically for $\chi$'s with soliton-like behaviors.  For
bare N and $\Delta$, all valence quarks are in $1S$ state, and so the
quark wave function is assumed to be
\begin{equation}
	q(x)
	=\frac{N_s}{\sqrt{4\pi}} \pmatrix{g(r) \cr
	i{\bf \sigma}\cdot\hat{r} f(r)}
	\xi_{s\mu} e^{-i\omega_s t},
\end{equation}
where $\xi_{s\mu}$ is the quark spin wave function.  As usual, the
scalar-isoscalar dielectric field $\chi$ is assumed to be
time-independent and spherically symmetric.

In the same spirit as the cloudy bag model, we treat the baryons as
composite, three quark systems, while simultaneously treating the
pions as an elementary, quantized fields.  As illustrated in
Figure~\ref{vertices}, the interaction hadron-space Hamiltonian
involving a single pion is then:
\begin{equation}
        H^B_{1\pi} =
        \sum_{\alpha,\beta}B_{\beta}^\dagger B_{\alpha}
        \int {d^3 k\, k_\mu \over (2\pi)^{3/2}}
         \left[ \vec{a}(k) \cdot
        \vec{\protect{\cal V}}_{\beta\alpha}^\mu({\bf k})
        + \mbox{h.c.}  \right],
\end{equation}
where
h.c. denotes Hermitian conjugate, $\vec{a}(k)$ is the pion
annihilation operator, and the vertex function $\vec{\protect{\cal
V}}_{\beta\alpha}^\mu$ is the matrix element of the quark pseudovector
current evaluated between transition baryon states:
\begin{eqnarray}
\vec{\protect{\cal V}}_{\beta\alpha}^\mu({\bf k})&=&
- {i\over 2f}  \int d^3x {e^{i{\bf k}\cdot {\bf x}}
\left\langle B_{\beta}\left|\overline q \gamma^\mu \gamma_5 \vec{\tau} q
\right| B_{\alpha}\right\rangle \over \sqrt{{(2 \pi)}^3 2 \omega_\pi(k)}}
 \\
&=& -{i\over 2f}{ u^{PV}(k) \left\langle B_{\beta}|{\bf \sigma}\cdot
\hat{\bf k} \vec{\tau} |B_{\alpha}\right\rangle^{sf}
 \over \sqrt{{(2 \pi)}^3 2 \omega_\pi(k)}}.\label{vertex}
\end{eqnarray}
The function $u^{PV}(k)$ in (\ref{vertex}) is the $\pi$NN pseudovector
form factor defined in the Appendix.  The braket in (\ref{vertex}) is
a spin-flavor matrix element which we reduce via the Wigner-Eckart
theorem:
\begin{eqnarray}
	\left\langle B_{\beta}|{\bf \sigma}\cdot\hat{\bf k} \vec{\tau}
	|B_{\alpha}\right\rangle^{sf} &=&
	-\sqrt{4\pi\over 3}\sum_M Y_{1M}^*(\hat{k})
	\lambda_{B_\alpha MB_\beta} \nonumber \\
	&& \times C_{\mu M \mu_{B_\beta}}^{s_B 1 s_{B_\beta}}
	C_{i_B i_M i_{B_\beta}}^{I_B I_M I_{B_\beta}},
\end{eqnarray}
where $C_{\mu M \mu_{B_\beta}}^{s_B 1 s_{B_\beta}}$ and $C_{i_B i_M
i_{B_\beta}}^{I_B I_M I_{B_\beta}}$ are Clebsch-Gordon coefficients
and the $\lambda_{B_\alpha MB_\beta}$ are the one-pion coupling
constants given in Table~\ref{cpl}.

The contact interaction  directly couples the initial pion and
baryon to the final pion and baryon, Figure~\ref{vertices}, and
takes the  form:
\begin{eqnarray}
        H^B_{2\pi} &=& \sum_{\alpha,\beta} B_{\beta}^{\dagger} B_{\alpha}
        \int {d^3k \,d^3k' \over (2\pi)^3}\nonumber\\
        &&\times \left[\left(\vec{a}^\dagger(k') \times k_\mu \vec{a}(k)\right)
        \cdot \vec{\protect{\cal W}}_{\beta\alpha}^\mu
        ({\bf k'},{\bf k})  + \mbox{h.c.} \right].
\end{eqnarray}
We evaluate the time and space components of the contact interaction
separately:
\begin{eqnarray}
	\vec{\protect{\cal W}}_{\beta\alpha}^T ({\bf k'},{\bf k}) &=&
	\frac{\omega_\pi(k)+\omega_\pi(k')} {4\pi^2
	f^2\sqrt{4\omega_\pi(k) \omega_\pi(k')}}
	\sum_{LM} Y_{LM}^* (\hat{k'}) Y_{LM}(\hat{k})\nonumber\\
	&&\times v_L^{CT}(k',k)
	\left\langle {B_{\beta}\left| -\frac{i}{2}\epsilon_{ijk}
	\tau_i \right|}B_{\alpha}\right\rangle^{sf} ,\\*[1ex]
	\vec{\protect{\cal W}}_{\beta\alpha}^S ({\bf k'},{\bf k})
	&=&
	\frac{1}{4\pi^2 f^2\sqrt{4\omega_\pi(k) \omega_\pi(k')}}
	 \sum_{JLMM'} Y_{LM}^* (\hat{k'}) Y_{LM}(\hat{k}) \nonumber\\
	&&\times C^{S'LJ}_{\mu' m'j} C^{SLJ}_{\mu MJ}
	 A^{JLS} v_L^{CS}(k',k)
	 \left\langle B_{\beta}\left|
	{-i\over 2\sqrt{6}}\epsilon_{ijk}\tau_i {\bf \sigma}
	\right|B_{\alpha}\right\rangle^{sf},\\*[1ex]
	A^{JLS} &=& -2\sqrt{6}\left\langle
	L\left\|{\bf L}\right\| L \right\rangle (-)^{J+S+L}
 	\left\{ \protect{\begin{array}{ccc} S'& S& 1\\ L &L &J
	\end{array}} \right\} . \label{ajls}
\end{eqnarray}
The spin-flavor matrix elements are
\begin{eqnarray}
	\left\langle {B_{\beta}\left| -\frac{i\epsilon_{ijk}}{2}\tau_i
	\right|}B_{\alpha}\right\rangle^{sf} &=&
	\sum_I \lambda_{\beta\alpha}^{TI}
	C_{i_B i_M i}^{I_B I_M I} C_{i_{B'} i_{M'} i}^{I_{B'} I_{M'} I},\\
	\left\langle {B_{\beta}\left|
	-{i\epsilon_{ijk}\over 2\sqrt{6}}\tau_i{\bf \sigma}
	\right|}B_{\alpha}\right\rangle^{sf}
  	&=&\sum_I \lambda_{\beta\alpha}^{SI}
	C_{i_B i_M i}^{I_B I_M I} C_{i_{B'} i_{M'} i}^{I_{B'} I_{M'} I},
\end{eqnarray}
where the two-pion coupling constants $\lambda_{\beta\alpha}^{TI}$ and
$\lambda_{\beta\alpha}^{SI}$ are given in Table~\ref{cpl'}.  In the
Appendix, the vertex functions $v_L^{CT}(k',k)$ and $v_L^{CS}(k',k)$
are related to the nucleon electromagnetic form factors.

\subsection{Vertex Functions with  Momentum-Projected Baryons}

The solutions of equations (\ref{field1})-(\ref{field2}) describe
quark orbits within baryons in which the COM may move. While this may
not be too serious a concern for bound-state spectra, it is for
scattering where the large momentum transfers make recoil effects
important.  We remove this spurious motion by a
Peierls-Yoccoz\cite{PY} projection to form an eigenstate of momentum
${\bf p}$:
\begin{equation}
	\left|B({\bf p})\right\rangle
	= {1\over N_B} \int d{\bf r} \, e^{i {\bf p} \cdot {\bf r}}\,
	q^\dagger_{\bf r} \,q^\dagger_{\bf r} \,q^\dagger_{\bf r}
	\left|C_{\bf r}\right\rangle ,
\end{equation}
where $N_B$ is a momentum-dependent normalization constant.
The operator $q^\dagger_{\bf r}$ creates a quark at point ${\bf r}$. The ket
$\left|C_{\bf r}\right\rangle$ represents the coherent state of the
dielectric field generated from the mean field solution\cite{Wilets}:
\begin{equation}\label{C}
	\left|C_{\bf r}\right\rangle
	= \exp \left[ \int d^3k \sqrt {\omega(k)/2}
	f_{\bf r}({\bf k}) a_{gb}^\dagger ({\bf k})\right]
	\left|0\right\rangle .
\end{equation}
In (\ref{C}), $\left|0\right\rangle $ is the vacuum state, $f_{\bf
r}({\bf k})$ is the Fourier transform of the scalar dielectric field
$\chi({\bf r})$, and $a_{gb}^\dagger ({\bf k})$ is the creation
operator for a scalar dielectric field quanta with energy
$\omega(k)=\sqrt{m_{gb}^2+ k^2}$.  Accordingly, the pseudovector
vertex function with recoil corrections is evaluated between the
momentum-projected baryon states in the Breit frame:
\begin{equation}
	\vec{\protect{\cal V}}_{\beta\alpha}^{\mu}({\bf k})
	={-i \left\langle B_{\beta}(-{{\bf k}\over 2})
	\right|\overline q(0)
	\gamma^\mu \gamma_5 \vec {\tau} q(0)
	\left|B_{\alpha}({{\bf k}\over 2})\right\rangle. \over 2f \sqrt{2
	\omega_\pi(k)}}  \label{V}
\end{equation}
After making these same projections and evaluations, the corresponding
contact interaction Hamiltonian (\ref{H}) takes the form:
\begin{equation}
	\vec{\protect{\cal W}}_{\beta\alpha}^\mu({\bf q}) =
	{\left\langle B_{\beta}(-{{\bf q}\over 2})\right|
	\overline q(0) \gamma^\mu \vec {\tau} q(0)
	\left|B_{\alpha}({{\bf q}\over 2})\right\rangle\over 4f^2
	\sqrt{4 \omega_\pi(k) \omega_\pi(k')}} ,
\end{equation}
where ${\bf q=k'-k}$ is the momentum transfer.  Detailed expressions
for the vertex functions are given in the Appendix.

The vertex function $\vec{\protect{\cal V}}^i_{\beta\alpha}(k)$ is of
some interest in its own right because it is related to the axial form
factor of the nucleon. In particular, ${\cal V}_{N,N}(k)$ in the $k
\rightarrow 0$ limit deteremines  $g_A$ of nucleon.  We find in our
calculations that $g_A$ calculated using momentum-projected states is
larger than that calculated statically.  Similar behavior is expected
for other meson-baryon coupling constants.
The form factor $\vec{\cal W}^\mu_{\beta\alpha}(q)$ for the contact
interaction is the matrix element of the
vector current, and has time and space components which are
proportional to the charge and magnetic form factors of the baryons.

\subsection{One Gluon Exchange Corrections}

While it is good that we have been able to solve the field equations
in the mean field approximation, the approximations involved leave the
nucleon and delta with the same mass. This degeneracy is removed
by including one gluon exchange corrections. Since the scalar
dielectric field $\chi$ is responsible for the confinement of gluons,
it is consistent to ignore the self-interaction of the residual
gluon. This means we drop the non-Abelian $gf_{abc}A_\mu^b A_\nu^c$
term in equation (\ref{F}) for the gluon field tensor $F^{a,\mu\nu}$.
Under the Coulomb gauge condition ${\bf \nabla}\cdot (\kappa {\bf A}^a) = 0$,
the equations for  the time and space components of the gluon fields become,
\begin{eqnarray}
	-{\bf \nabla} \cdot (\kappa {\bf \nabla} A^{a,0})
	&=& j^{a,0} \\
	\kappa \partial_0^2 \vec A^a
	- {\bf \nabla}^2 (\kappa {\bf A}^a)
	&+& {\bf \nabla}
	\times (\kappa {\bf A}^a \times {\bf \nabla} \ln \kappa)
	= {\bf j}^a_t
\end{eqnarray}
Here $j^{a,\nu} = {1\over 2} g \overline q\gamma^\nu
\lambda^a q$ is the quark  current, and
${\bf j}^a_t$ is its transverse component.
% $K(r)$ is the
%dielectric coefficient (\ref{K}), and $\lambda^a$ is a Gell-Mann
%matrix.
We follow Bickeboller et al.\cite{BGW} and solve these
equations for $A^a_\mu$ with Green's function techniques.  As is done in bag
model calculations\cite{bag}, the contribution of the gluon electric
energy is neglected. The gluon magnetic energy can then be expressed in
the form:
\begin{equation}
	E_{mag} = \left\langle B(0){\left|
	\sum_{a, i< j} {\bf A}^a(i)
	\cdot  {\bf j}^a(j)\right|B(0)}\right\rangle,
\end{equation}
where ${\bf A}^a(i)$ is the gluon field due to i-th quark.
The bare baryon mass is the sum of the expectation values of the bare
Hamiltonian for a zero-momentum baryon state plus the gluon magnetic
energy:
\begin{eqnarray}
        m_{B_0}^{(0)} &=& \left\langle B(0){\left|H_{bare}
        \right|B(0)}\right\rangle  + E_{mag},\\
	H_{bare} &=&
	\int d^3 x \left[ \sum_i q^\dagger_i
	\left(-i{\bf \alpha} \cdot {\bf \nabla}
	+ {\gamma^0 m_q\over \chi}\right)q_i \right. \nonumber \\
	&& \left. + {\sigma_v^2 \over 2}\left\{({\bf \nabla} \chi)^2
	+ \pi^2\right\} +U(\chi) \right],
\end{eqnarray}
where $\pi$ is the conjugate momentum of the $\chi$ field.
When these steps are followed, the masses of the bare nucleon and
delta can be written as
\begin{eqnarray}
        m_N^{(0)} &=& m_{gb}(C_1-\alpha_s C_2), \\
        m_\Delta^{(0)} &=& m_{gb}(C_1+\alpha_s C_2),
\end{eqnarray}
where $C_1$ and $C_2$ are  constants determined from the
mean field solutions.

\section{Unitary}

In the present work we concentrate on $\pi$N scattering in the energy
region from threshold through the $\Delta$ resonance.  Even though the
$\pi \Delta$ channel is closed for these energies, we include coupling
to an explicit $\pi \Delta$ channel in order to include its effect on
$\pi$N scattering and to permit a subsequent extension of the theory
to $\Delta$ production. To ensure unitarity we iterate the tree-level
diagrams using the relativistic Lippmann-Schwinger equation (\ref{T}).
The resulting T matrices have the desired renormalized poles together
with the pion-dressed vertex functions.  In Figure~\ref{fig.pots} we
show the three distinct terms which contribute in lowest order to the
$\pi$N and $\pi\Delta$ potentials:\\ 1. {\bf Direct Born term},
Figure~\ref{fig.pots}a.  This part of the potential arises from two
elementary vertices connected by an intermediate bare baryon state
$B_0$. In the partial-wave basis,
\begin{eqnarray}
	\lefteqn{V_{\beta \alpha}^{(a)LJI}(k',k) =
	{\delta_{I I_{B_0}} \delta_{J J_{B_0}} \delta_{L 1} \, k^{'} \, k
	\over 48 \pi f^2 \sqrt{4\omega_\pi (k)\omega_\pi (k')}} }
	\nonumber \\
	&& \times \sum_{B_0} \lambda_{B'M'B_0} \lambda_{BMB_0}
	{u^{PV}(k') u^{PV}(k) \over E-m_{B_0}^{(0)} }.
\end{eqnarray}
Here $E$ is the COM energy and $m_{B_0}^{(0)}$ is the bare mass of the
intermediate nucleon or delta. Spin and isospin conservation
restricts the intermediate state to the $P_{11}$ and $P_{33}$ channels.\\
2. {\bf Crossed Born term}, Figure~\ref{fig.pots}b. This term
arises from crossing the external pion lines of the pole
diagram. Since the intermediate state contains two pions and one
baryon, this modifies the propagator and so:
\begin{eqnarray}
	\lefteqn{V_{\beta \alpha}^{(b)LJI}(k',k) ={\delta_{L 1} \, k' \, k
	\over 48\pi f^2 \sqrt{4\omega_\pi(k)\omega_\pi(k')}}} \nonumber\\
	&&\times \sum_{B_0} \lambda_{B_0 M B'} \lambda_{B_0 M' B}
	{\eta_{\beta\alpha}^{LJI} u^{PV}(k') u^{PV}(k)
	\over E-m_{B_0}^{(0)}-\omega_\pi (k)-\omega_\pi(k')}.
\end{eqnarray}
The spin-isospin coefficient is defined as
\begin{eqnarray}
	\eta_{\beta\alpha}^{LJI} &=& (-1)^{S_B+S_{B'}+I_B+I_{B'}+I_M+I_{M'}}
	\hat{S_B}\hat{S_{B'}}\hat{I_B}\hat{I_{B'}} \nonumber \\
	&& \times \left\{ \protect{\begin{array}{ccc}I_M & I_B & I\\ I_{M'}
	& I_{B'} & I_{B_0} \end{array}} \right\}
	\left\{ \protect{\begin{array}{ccc}S_B & 1 & J\\ S_{B'} & 1 & S_{B_0}
	\end{array}} \right\}.
\end{eqnarray}
Recoil of the intermediate baryons are ignored in our calculations,
and thus the crossed term contributes only to P-wave $\pi$N
scattering.\\ 3. {\bf Contact term,} Figure~\ref{fig.pots}c. The
contact term arises from $H_{2\pi}$, the two-pion vertex of the
interaction Hamiltonian.  This term gives rise to two potentials,
$V_{\beta\alpha}^{CT}$ due to the time component, and
$V_{\beta\alpha}^{CS}$ due to the space component. In the partial-wave
basis,
\begin{eqnarray}
	V_{\beta \alpha}^{(CT)LJI}(k',k)&=&
	{(\omega_\pi(k)+\omega_\pi(k'))\lambda_{\beta\alpha}^{TI}
	v_L^{CT}(k',k) 	\over
	8\pi f^2 \sqrt{4\omega_\pi(k)\omega_\pi(k')}}, \\
	V_{\beta \alpha}^{(CS)LJI}(k',k)&=&
	{\lambda_{\beta\alpha}^{SI} A^{JLS} v_L^{CS}(k',k)
	\over 8\pi f^2 \sqrt{4\omega_\pi(k)\omega_\pi(k')}},
\end{eqnarray}
where $A^{JLS}$ has been defined previously (\ref{ajls}), and the spin-flavor
coupling constants are given in Table~\ref{cpl'}.

\section{Results}

We have used a CDM to generate the potential terms for the
relativistic Lippmann-Schwinger equation (\ref{T}).  The pion
interaction resulting from the iteration of the potential by the
Lippmann-Schwinger equation ensures two body unitary of $T$ and
renormalizes the $\pi$BB' vertex and the bare baryon masses.  The
energies of the poles of T are identified with renormalized masses for
the physical states, the residues of the poles are identified with
renormalized coupling constants, and the momentum distributions around
the poles are related to the form factors.

When fitting data we always start off with a value for the glueball
mass $m_{gb}\simeq$ 1 GeV. The precise value does not appear to be
important, but its general magnitude does set the energy scale in the
equations of motion.  We keep the shape parameter of the
self-interaction of dielectric field fixed at $\alpha=24$ since
previous studies with the pseudoscalar version of the CDM\cite{CDM1}
have shown that the static properties of baryons are insensitive to
$\alpha$ for this size of $m_{gb}$. Each time we obtain a solution of
the field equations, we adjust the parameters $m_{gb}$, $\alpha_s$,
and $f$ to best describe the $\pi$N phase shifts and the masses of the
nucleon and delta.  Somewhat surprisingly, we found that we can
reproduce the experimental masses for a large range of quark masses
and bag constants just by adjusting the values for $m_{gb}$ and
$\alpha_s$. Specifically, we find that we can vary the bag constant
$B^{1/4}$ between 100 MeV and 150 MeV, and the quark mass between 40
MeV and 120 MeV. While 40 MeV may appear small, the dielectric field
also becomes small near the origin (the ``one phase'' solution in
which the wave function has no abrupt phase transition).
Consequently, the 40 MeV quark has an effective mass inside of a
nucleon of $m_q/\chi(r\simeq 0) \simeq$ 100 MeV.

The fitted parameters and deduced bare masses are given in
Table~\ref{tab0}. The value $f=93$\,MeV for the pion decay constant is
the accepted value\cite{books}, and the value $f=90$\,MeV is very
close. Our value $\alpha_s \simeq 0.12$ for the quark-gluon coupling
constant is much smaller than the MIT bag result, $\alpha_s \simeq 2.2$
\cite{MIT} or the favored value in the Friedberg-Lee model.
In fact Bickeboller et al.\cite{BGW} have found that $\alpha_s$ can vary over
two orders of magnitude depending  on the choice of the dielectric coefficient
and the parameters of the self-interaction of the dielectric field.
In our case, a small value of $\alpha_s$ compensates for our model of
the dielectric field, $\kappa(\chi) = \chi^4$,
which otherwise would produce too large a color-magnetic energy.
In addition, the pion-dressing in our model makes up for part
of the  $N$ and $\Delta$ mass difference.
This also makes the value of $\alpha_s$,
which measures the strength of OGE in our model,
smaller than in those models without dressing.
Our fitted value $m_{gb} \simeq
1153$\,MeV for the glueball mass is consistent with the values found
in other models\cite{Wilets}, and is in the region where glueball may
occur\cite{booklet}.  Finally, we see in
Table~\ref{tab0} that the effect of renormalization on the nucleon and
delta masses is to move the bare masses down by $\sim 250$\,MeV. Our
calculation of scattering lengths indicates the need for some
additional repulsion in our model, and we suspect that that repulsion
would lead to less mass renormalization.

\subsection{Coupling Constants, Masses, and Form Factors}

As indicated in (\ref{H}), we have deduced the CDM Hamiltonians
$H_{1\pi}$ and $H_{2\pi}$ from an expansion of the Lagrangian in
inverse powers of the pion's weak decay constant $f$, or equivalently,
as a series expansion in the $\pi$NN coupling constant $f_{\pi NN}$. This
equivalence follows from the Goldberger-Trieman relation\cite{cbm}:
\begin{equation}        \label{GT}
        \frac{1}{2f} =  \sqrt{4\pi}\, \frac{f_{\pi NN}}{m_{\pi}}.
\end{equation}
While the relation (\ref{GT}) relates  the bare coupling constants,
our study also produces renormalized
coupling constants. We extract the bare coupling constant
$f^{(0)}_{\pi NN}$ from our final $T$ by comparing the vertex function
computed with the CDM to that of standard Chew-Low
theory\cite{cbm}:
\begin{equation}
        v_{i}(k) = i \sqrt{4\pi \over 2 \omega_k}
        {f^{(0)}_{\pi NN} \over m_{\pi}} g(k) \tau_i {\bf \sigma \cdot k}.
\end{equation}
We then extract the renormalized $\pi$NN coupling constant
$f_{\pi NN}$ and the renormalized $\pi$N$\Delta$ coupling
constant $f_{\pi N\Delta}$ by making Laurent expansions of the
computed T matrix around its poles:
\begin{eqnarray}
        \hat{T}(k',k;E\simeq m_N) &\simeq& f^2_{\pi NN}
        \frac{g_{\pi NN}(k') g_{\pi NN} (k)}{E-m_N} +
        \cdots, \label{pole1}\\ \hat{T}(k',k;E\simeq m_\Delta)
        &\simeq& f^2_{\pi N\Delta} \frac{g_{\pi N\Delta}(k') g_{\pi
         N\Delta} (k)}{E-m_\Delta} + \cdots \ .  \label{pole2}
\end{eqnarray}
This expansion permits us to deduce not only coupling constants  but
also the $\pi$NN  and $\pi$N$\Delta$ form factors $g_{\pi NN} (k)$ and
$g_{\pi N\Delta}(k)$ from the momentum dependence of the T matrices in
(\ref{pole1}) and (\ref{pole2}). It is particularly interesting to
compare these form factors with those calculated in other approaches
since the form factors are related to the CDM quark wave functions
(which should be fairly realistic), and since they are affected by the
renormalization process.

While we have already indicated that fits to the data do not determine
a unique set of parameters, the different parameter sets are
correlated. In the upper part of Figure~\ref{fvsf} we show the
relation between the strong coupling constant $f_{\pi NN}$
deduced from our fits and the bare, weak decay constant $f$ (note the
false origin). The dashed curve relates the bare values of
$f_{\pi NN}$ as determined from the Goldberger-Trieman relation
(\ref{GT}), or equivalently, from the potential directly. The solid
curve relates the renormalized values determined via (\ref{pole1}). We
see that there is a similar functional dependence in each case with an
$\sim 15\%$ renormalization effect.

Although not shown in the figure, we have also calculated
$f_{\pi N\Delta}$ from the residue of the $\Delta$ pole and
find:
\begin{equation}
        {f_{\pi N\Delta} \over f_{\pi NN}} \simeq 1.87.
\end{equation}
This ratio is quite close to the experimental value of $\sim 2$,
closer in fact than the SU(6) prediction\cite{cbm} of $6\sqrt{2}/5
\simeq 1.70$. This  clearly shows the importance of renormalization.

In the lower part of Figure~\ref{fig.g(k)} we show the $\pi$NN form
factors deduced from the behavior of the T matrix near the nucleon
pole. The two different curves correspond to two different values of
the pion decay constant $f$.  One can estimate from the falloff that
the range $R \simeq 0.5$fm.  To compare with other models which have a
monopole form factor, we assume
\begin{eqnarray}
        g(k\rightarrow 0) &\simeq& {1 \over 1+ k^2R^2} ,
	\label{mono}\\
        \Rightarrow R = \lim_{k \rightarrow 0}
        \sqrt{g'(k)\over -2k}
        &=& \cases{0.47 \,\mbox{fm} & $f=90$ MeV, \cr 0.48 \,\mbox{fm}
        &$f=93$ MeV. \cr}
\end{eqnarray}
This $R$ is comparable with the value $R \simeq
R_{\mbox{bag}}/\sqrt{10}$ obtained from the cloudy bag model with a
bag radius $R_{\mbox{bag}} \simeq 1$ fm. The meson exchange models,
however, use much smaller $R \sim 0.15 $ fm which produces a much
harder form factor.\footnote{We remind the reader that the monopole
approximation (\ref{mono}) is for comparison only and that this radius
is determined from the behavior of the form factor in the $k
\rightarrow 0$ limit.} In practice, a monopole form factor falls off
much slower than the form factor calculated in our model. If we
replace our form factor with the monopole form factor defined above
and use it in an actual T-matrix calculation, we would get poor
agreement with the experimental phase shifts. Thus, in contrast to the
meson exchange models, pion-nucleon scattering seems to require soft
form factors like the ones obtained in our model. A similar conclusion
has also been reached recently by others\cite{TSHL}.

\subsection{S and P Wave Phase Shifts}

In Figure~\ref{fig.del} we compare the low-energy S- and P-wave phase
shifts calculated with the CDM model to the experimental phases
obtained from the {\em said} analysis\cite{said}.  The solid and
dashed curves are for parameter sets with $f$=90 MeV and $f$=93 MeV
respectively. In Table~\ref{tab1} a comparison of just the scattering
lengths is given. We see excellent agreement in the $P_{33}$ channel,
with the energy of the delta and its width reproduced well. The small
P-wave phases are always more of a challenge. The scattering volumes
for the $P_{31}$ and $P_{13}$ channels are predicted well, with the
calculated energy dependence in the $P_{31}$ phase very similar to the
data.  In Table~\ref{tab1} we give values for the scattering lengths
calculated in Born approximation, that is, when only the potential
term in the Lippmann-Schwinger equation (\ref{T}) is used for $T$. As
expected, these values agree with the chiral limits:
\begin{equation}
        a_1= {\mu \over 4 \pi f^2},
        \ \ \ a_3= -{\mu \over 8 \pi f^2},
\end{equation}
where $\mu$ is the $\pi$N reduced mass.  Comparison of the
Born results to experiment and to our full calculation indicates that
renormalization is large, and that some additional repulsion is
needed.

The $P_{11}$ channel is a particular problem for simple models like
ours. Increasing the agreement with the $P_{33}$ phases decreases the
agreement with the $P_{11}$ phases.  We have assumed that the
Roper(1440) is a $(1s)^2(2s)$ three-quark state, and included it as an
intermediate state for both s- and u-channel diagrams.  The results
show that the elementary Roper does provide the attraction needed for
the phase shift to change sign from negative to positive, but does not
do much to improve the fit at lower energies. In addition, we find
that the $P_{11}$ amplitude does not have the desired pole at the
physical Roper resonance unless we introduce extra, free parameters.
Indeed, the calculations of Elsey and Afnan\cite{afnan1} and Pearce
and Afnan\cite{afnan2} indicate that the coupling of three-body
($\pi\pi$N) channels is important, as are some $1/f^4$ diagrams not
present in our calculation. These should be included in the
continuation of the present work.

As is true for the small P-wave phases, the predicted S-wave phase shift
shown in Figure~\ref{fig.del} show the need for additional repulsion.
Cooper and Jennings\cite{coop} indicate that agreement may require
the inclusion of certain $1/f^4$ graphs which cancel at $m_{\pi}=0$
but produce repulsion for $m_{\pi} > 0$.

\section{Summary and Conclusion}

We have studied low-energy $\pi$N elastic scattering in a chiral
version of the color dielectric model. Our calculations are similar to
those done with the cloudy bag model\cite{triumf}. Our major
improvement is a more natural and realistic inclusion of quark
confinement, the inclusion of baryon-recoil effects, and the coupling
of the $\pi$N and $\pi\Delta$ channels.  The CDM is used to derive
effective potentials containing terms up to second order in the
$\pi$NN coupling, and these potentials are used as input to
relativistic, coupled-channel Lippmann-Schwinger equations. The model
reproduces the physical nucleon and delta masses, and then predicts
the $\pi$NN and $\pi$N$\Delta$ coupling constants and form factors,
and low-energy $\pi$N scattering.

We find much success for such a simple model. Specifically, we can
simultaneously obtain the correct renormalized masses for the nucleon
and delta, the correct delta width, the $\pi$NN coupling constant to
within 5\% of it experimental value, the ratio of the renormalized
$\pi$N$\Delta$ to $\pi$NN coupling constants which is better than the
SU(6) value, excellent agreement with the $P_{33}$ phase shifts from
threshold through the delta resonance energy, good agreement with the
$P_{13}$ and $P_{31}$ scattering volumes, and good agreement with the
energy dependence of the $P_{31}$ phase shift for the first 300 MeV of
kinetic energy.  Better agreement with the S-wave and small P-wave
phases requires additional repulsion in the model, possibly obtained
by one-loop corrections. The model could also be improved by obtaining
solutions to the field equations beyond those of the mean field
approximation and by the inclusion of additional dielectric
fields\cite{pirner}.

\appendix
\section*{Form Factors for $\pi$N Scattering}

We give here the complete form factors for the ($\pi$N,$\pi\Delta$)
system. Since the $\Delta$ differs from the $N$ only in spin-isospin
orientations, they have the same space wave function and thus the same
form factors.  The pseudovector vertex functions with the momentum
projected states evaluated in the Breit frame are:
\begin{eqnarray}
   \vec{\protect{\cal V}}_{\beta\alpha}^{\mu}({\bf k})
    &=&-{i \left\langle B_{\beta}(-{{\bf k}\over 2})\right|\overline q(0)
    \gamma^\mu  \gamma_5 \vec{\tau} q(0)
	\left|B_{\alpha}({{\bf k}\over 2})\right\rangle
	\over 2f \sqrt{2\omega_\pi(k)}} \nonumber\\
	&=& {
	-i\int d{\bf x} d{\bf y} \, e^{-i{\bf k\cdot (x + y)/2}}
  \mbox{}_{\bf y}\left\langle B_{\beta}|
  \overline q(0) \gamma^\mu \gamma_5 \vec{\tau} q(0)
  |B_{\alpha}\right\rangle_{\bf x}
 \over
  2f \sqrt{2 \omega_\pi(k)} \, N_{B_\beta} N_{B_\alpha} }\nonumber \\
  &=&
{i \over 2f \sqrt{2\omega_\pi(k)}} \,u^{PV}(k)\left\langle B_\beta\left|
{\bf \sigma}\cdot \hat{\bf k}\, \vec{\tau}\right| B_\alpha \right\rangle^{sf}.
\end{eqnarray}
Where $N_{B_\alpha} =
\sqrt{
%{\left\langle B_\beta(-{{\bf k}\over 2}) | B_\beta(-{{\bf k}\over 2})
%\right\rangle}\,
 {\left\langle B_\alpha({{\bf k}\over 2}) | B_\alpha({{\bf k}\over 2})
\right\rangle }}$ is a momentum-dependent normalization constant for
a projected baryon state,
 and $ \left|B_\alpha\right\rangle_{\bf x}
        = q^\dagger_{\bf x} \,q^\dagger_{\bf x} \,q^\dagger_{\bf x}
        \left|C_{\bf x}\right\rangle$ is a localized baryon state.
 The pseudovector form factor  has the form, $u^{PV}(k)=N(k)/D(k)$, with
\begin{eqnarray}
  N(k)&=& \int z^2 dz N_q^2(z) N_\chi(z) \int d^3 r \nonumber\\
 	&&\times  \left\{ g_{+} g_{-} j_0(kr)
  	-{f_{-} f_{+} \over 3r_{-} r_{+}} \left[[j_0(kr)+4j_2(kr)] r^2
  	-[j_0(kr)+4j_2(kr)P_2(\hat{r}\cdot\hat{z})]{z^2\over 4}\right]
	 \right\}, \\
  D(k) &=& \int z^2 dz j_0(kz/2) N_q^3 (z) N_\chi (z), \\
  N_q(z) &=& \int d^3 r \left[g_{+}g_{-}+
 	{f_{+}f_{-} \over r_{+} r_{-}} \left(r^2 - {z^2 \over 4}\right)
	\right], \\
  N_\chi(z) &=& \exp\left[2\pi\int k^2 dk \,\omega(k)\left| f_0(k) \right|^2
	 j_0(kz)\right].
\end{eqnarray}
In the above equations, we have defined $r_\pm=\left|{\bf r}\pm
{\bf z}/ 2\right|$, $g_\pm=g(r_\pm)$, and $f_\pm=f(r_\pm)$. The
function $f_0(k)$ is the Fourier transform of the mean field solution
of the classical dielectric field.  Without the COM correction,
$g_\pm=g(r)$, $f_\pm=f(r)$, and the $\pi$NN pseudovector form factor
is reduced to the static limit:
\begin{eqnarray}
	u^{PV}(k)&=&N_s^2 \int r^2 dr \left\{ \left[g(r)^2
	- {f^2(r) \over 3}\right] j_0(kr)\right.\nonumber \\
	&& \left. -{4f^2(r) \over 3} j_2(kr) \right\}.
\end{eqnarray}

The vertex functions for the contact interaction   are
\begin{eqnarray}
 \vec{\protect{\cal W}}_{\beta\alpha}^\mu({\bf q}) &=&
  	{1\over 8f^2 \sqrt{ \omega_\pi(k) \omega_\pi(k')} }
 	\left\langle B_{\beta}(-{{\bf q}\over 2})\right|\overline q(0)
	\gamma^\mu \vec{\tau} q(0)
     	\left|B_{\alpha}({{\bf q}\over 2})\right\rangle \nonumber \\
	&=& {1 \over 8f^2 \sqrt{ \omega_\pi(k) \omega_\pi(k')}
 	\, N_{B_\beta} N_{B_\alpha} }\nonumber \\
	&&\times   \int d{\bf x}\,d{\bf y}\, e^{-i{\bf q} \cdot
	({\bf x + y})/2} {_{\bf y}\!\left\langle B_{\beta}\right|
	\overline q(0) \gamma^\mu \vec{\tau} q(0)
 	\left|B_{\alpha}\right\rangle_{\bf x}}
%{\sqrt{ \left\langle B_{\beta}(-{{\bf q}\over 2})\right|
%\left. B_{\beta}(-{{\bf q}\over 2})\right\rangle \left\langle
%B_{\alpha}({{\bf q}\over 2})\right|
%\left. B_{\alpha}({{\bf q}\over 2})\right\rangle }
 \nonumber \\
	&=& {1 \over 8f^2 \sqrt{ \omega_\pi(k) \omega_\pi(k')}}\nonumber \\
	&&\times \left\{ \begin{array}{ll}
	u^{CT}(q) \left\langle B_{\beta}| \vec{\tau}
	|B_{\alpha}\right\rangle^{sf}, & \mbox{time}, \\
	u^{CS}(q) \left\langle B_{\beta}| -{\bf \sigma}\times \hat{\bf q}
	\, \vec{\tau} |B_{\alpha}\right\rangle^{sf},
	& \mbox{space}.\end{array} \right.
\end{eqnarray}

The form factors for the time and space components  are respectively:
\begin{eqnarray}
	u^{CT}(q) &=& {N^{CT}(q)\over D(q)}, \ \ \ \
	u^{CS}(q) = {N^{CS}(q)\over D(q)}, \\
	N^{CT}(q) &=& N_s^2\int z^2dz N_q^2(z)N_\chi(z)
	\int d^3\, r j_0(qr)\nonumber\\
	&&\times \left[g_+ g_- + \left(r^2 -{z^2\over 4}\right)
	{f_{+} f_{-} \over r_{+} r_{-}}\right], \\
	N^{CS}(q) &=& N_s^2\int z^2dz N_q^2(z)N_\chi(z)
	\int d^3\, r {j_1(qr)\over qr}\nonumber\\
	&&\times \left[r^2 \left({g_+ f_-\over r_-} + {g_-f_+\over r_+}\right)
	- {\hat{r}\cdot\hat{z}\over 2} \left({g_+f_-\over r_-}-{g_-f_+\over
	r_+}\right)\right].
\end{eqnarray}
Their corresponding static expressions are:
\begin{eqnarray}
u^{CT}(q)&=&\int  dr\, r^2 j_0(qr) \left[g^2(r)+f^2(r)\right],\\
u^{CS}(q)&=&\int dr \, r^2 {j_1(qr)\over q} g(r)f(r).
\end{eqnarray}
Finally, the relationships between the electromagnetic form factors
and  the partial-wave decomposed radial potentials are:
\begin{eqnarray}
	v_L^{CT}(k',k)&=&\int_{-1}^{1} dx P_L(x) \, u^{CT}(q), \\
	v_0^{CS}(k',k)&=& 0, \\
	v_1^{CS}(k',k)
	&=& {k'k\over 6}\int_{-1}^{1} dx \,[1-P_2(x)] \, u_{CS}(q).
\end{eqnarray}

%\widetext

\begin{table}
\caption{One-pion spin-flavor coupling constants.} \label{cpl}
\begin{tabular}{ccc}
 & $\lambda_{BMB_0}$& \\ \hline
&$\pi$N & $\pi\Delta$\\ \hline
N & 5 & $4\sqrt{2}$ \\
$\Delta$& $2\sqrt{2}$ & 5
\end{tabular}\end{table}

\begin{table}
\caption{Two-pion spin-flavor coupling constants.} \label{cpl'}
\begin{tabular}{ccccccccc}
 & \multicolumn{4}{c}{$\lambda_{\beta\alpha}^{TI}$}       &
		     \multicolumn{4}{c}{$\lambda_{\beta\alpha}^{SI}$}  \\ \hline
 & \multicolumn{2}{c}{$I={1\over 2}$} &\multicolumn{2}{c}{$I={3\over 2}$}&
\multicolumn{2}{c}{$I={1\over 2}$} &\multicolumn{2}{c}{$I={3\over 2}$}
\\
 & $\pi$N & $\pi\Delta$ & $\pi$N & $\pi\Delta$ &
 $\pi$N & $\pi\Delta$ & $\pi$N & $\pi\Delta$\\ \hline
 $\pi$N&   -1 & 0 & ${1\over 2}$ & 0 &
 -${5\over 3}$ & -${2\sqrt{2}\over 3}$ & ${5\over 6}$ & $-{2\sqrt{5}\over 3}$
\\
$\pi\Delta$& 0 & $-{5\over 2}$ & 0 & -1 &
  -${2\sqrt{2}\over 3}$ & ${5\sqrt{10}\over 6}$ &
  -${2\sqrt{5}\over 3}$ & ${\sqrt{10}\over 3}$
\end{tabular}\end{table}

\begin{table}
\caption{Fitted parameters and deduced bare masses for the CDM.} \label{tab0}
\begin{tabular}{ccccc}
 $f$ (MeV) & $\alpha_s$ & $m_{gb}$ (MeV) & $m_N^{(0)}$ (MeV)&
	$m_{\Delta}^{(0)}$ (MeV)\\ \hline
90 & 0.112 & 1178 & 1239 & 1489\\
93 & 0.126 & 1127 & 1171 & 1439
\end{tabular}\end{table}

\begin{table}
\caption{S and P wave scattering lengths in fm where
$(k a_{\alpha})^{2L+1} =
\tan\delta_{\alpha}$.} \label{tab1}
\begin{tabular}{cccccc}
$L_{2I\,2J}$ & $a_{CDM}$ & $a_{CDM}$&
$a_{Born}$ & $a_{CBM}$ &Exp.\\
 & ($f$=90\,MeV) &($f$=93\,MeV)& & &\\
\hline
$S_{11}$  & +0.398	&   +0.342& 0.237	&+0.42    & +0.243\\
$S_{31}$ & -0.098	 & -0.093& -0.119	&-0.07  &  -0.130\\
$P_{11}$ & -0.892  	&  -0.860& 0.954 	& &   -0.569\\
$P_{31}$ & -0.412 	 & -0.420& -0.426 	& &  -0.483\\
$P_{13}$ & -0.355   	&  -0.366& -0.357	& &  -0.375\\
$P_{33}$ & +0.713	  & +0.712& 0.583  	& &  +0.839\\
\end{tabular}\end{table}

%Figure captions%=========================================

\begin{figure}
\caption{The linear and quadratic term in $1/f$ contributing
to  the interaction Hamiltonian.} \label{vertices}
\end{figure}

\begin{figure}
\caption{The three lowest-order terms of the potential
contributing to $\pi$N scattering.} \label{fig.pots}
\end{figure}

\begin{figure}
\caption{({\em Top}) The relation between the $\pi$NN  coupling
	constant and the pion decay constant $f$. ({\em Bottom}) The
	 $\pi$NN form factor for two different values of the pion
	decay constant $f$.} \label{fvsf}
\end{figure}

\begin{figure}
\caption{The $\pi$NN  form factors
deduced from the behavior of the T matrix near the nucleon pole.}
\label{fig.g(k)}
\end{figure}

\begin{figure}
\caption{The S- and P-wave $\pi$N phase shifts. The solid and
	dashed curves are for parameter sets with $f$=90 and $f$=93
	respectively, and the dots are the experimental phase
	shifts.}\label{fig.del}
\end{figure}

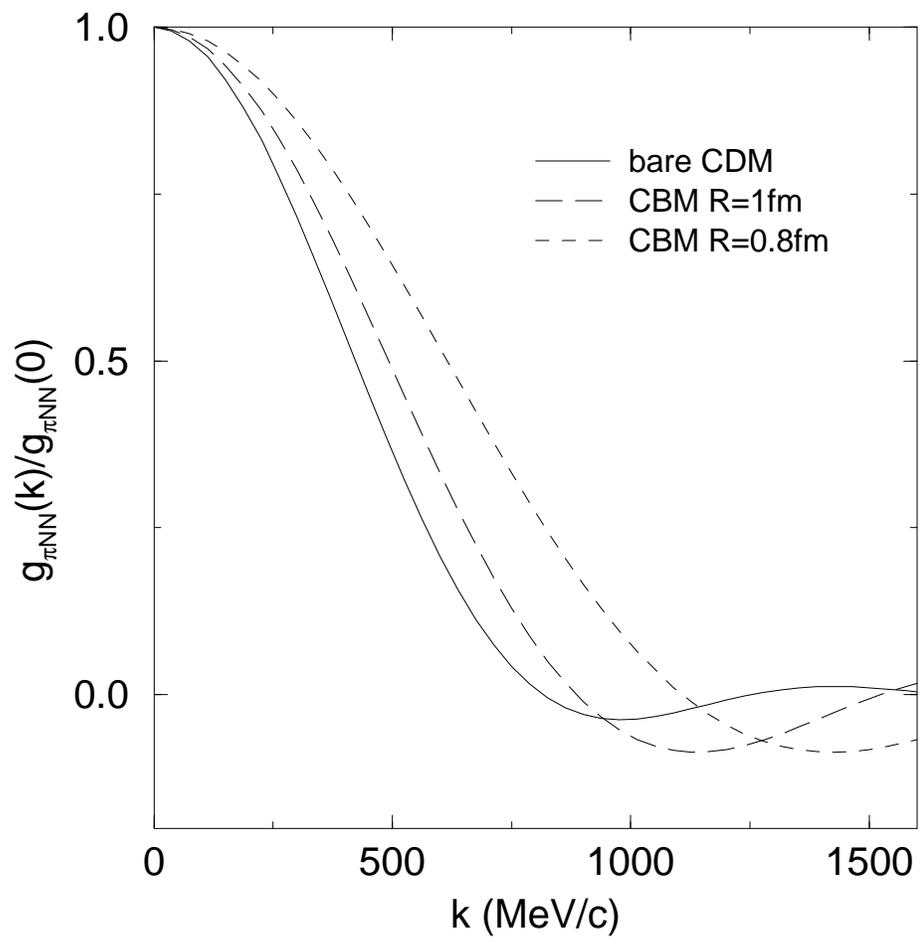
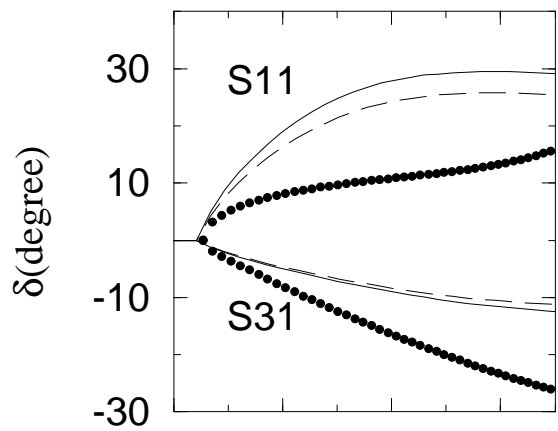
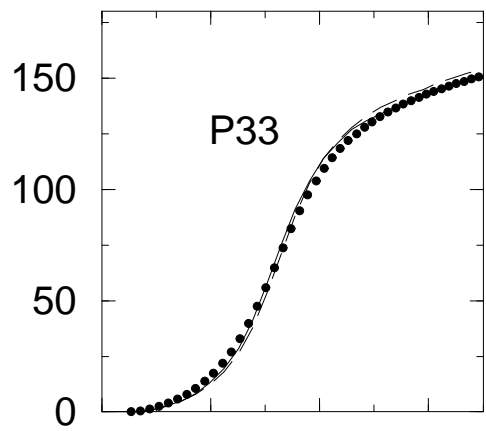
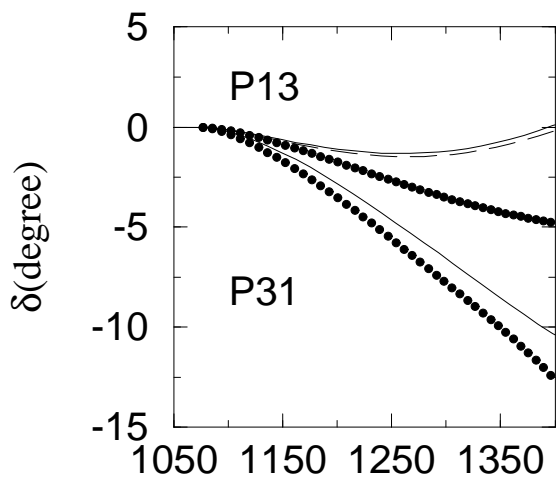
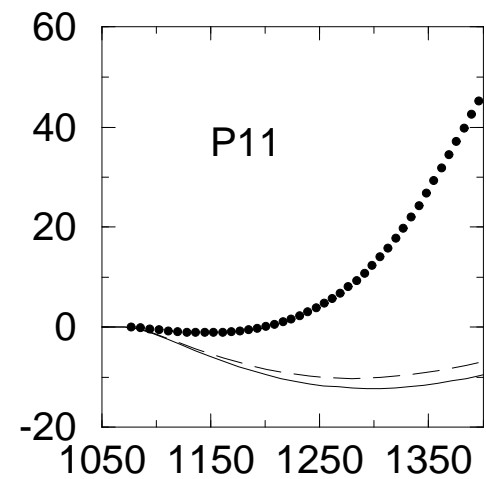
\end{document}